\documentclass[12pt]{article}
\usepackage{multirow}
\usepackage{booktabs}
\usepackage{mathrsfs}
\usepackage{fullpage}
\usepackage{amsmath}
\usepackage{graphicx}%
\usepackage{amsfonts}%
\usepackage{amssymb}
\usepackage{afterpage}
\usepackage{xurl}
\usepackage{fancyhdr}
\usepackage{bm}
\usepackage{isomath}
\usepackage{amsmath}
\usepackage{amsbsy}
\usepackage{amssymb}
\usepackage{amscd}
\usepackage{amsfonts}
\usepackage{graphicx}

\usepackage{verbatim}
\usepackage{euscript}
\usepackage{alltt}
\usepackage{stmaryrd}
\usepackage{float}
\usepackage[titletoc,title]{appendix}
\usepackage{caption}
\usepackage[english]{babel}
\usepackage{enumitem}
\usepackage{subcaption} 
\usepackage{layout}
\usepackage{longtable}
\usepackage{tikz}

\usepackage[linktoc=all]{hyperref}
\hypersetup{colorlinks=true,
	linkcolor=blue,
	citecolor=blue
}

\usepackage[margin=1in]{geometry}
\usepackage{titlesec}

\titleformat{\paragraph}
{\normalfont\normalsize\bfseries}{\theparagraph}{1em}{}
\titlespacing*{\paragraph}
{0pt}{3.25ex plus 1ex minus .2ex}{1.5ex plus .2ex}

\PassOptionsToPackage{normalem}{ulem}
\usepackage{ulem}

\newcommand\blfootnote[1]{%
  \begingroup
  \renewcommand\thefootnote{}\footnote{#1}%
  \addtocounter{footnote}{-1}%
  \endgroup
}

\begin{document}
	
\pagenumbering{roman}

\title{PhySRNet: Physics informed super-resolution network for application in computational solid mechanics}



\author{Rajat Arora\thanks{Senior member of technical staff at Advanced Micro Devices, Inc. (AMD). rajat.arora9464@gmail.com.}}

\date{}

\maketitle

\begin{abstract}



\noindent  Traditional approaches based on finite element analyses have been successfully used to predict the macro-scale behavior of  heterogeneous materials (composites, multicomponent alloys, and polycrystals) widely used in industrial applications.  However, this necessitates the mesh size to be smaller than the characteristic length scale of the microstructural heterogeneities in the material leading to computationally expensive and time-consuming calculations. The recent advances in deep learning based image super-resolution (SR) algorithms open up a promising avenue to tackle this computational challenge by enabling researchers to enhance the spatio-temporal resolution of data obtained from coarse mesh simulations. However, technical challenges still remain in developing a high-fidelity SR model for application to computational solid mechanics,  especially for materials undergoing large deformation. This work aims at developing a physics-informed deep learning based super-resolution framework (PhySRNet) which enables reconstruction of high-resolution deformation fields (displacement and stress) from their low-resolution counterparts without requiring high-resolution labeled data. We design a synthetic case study to illustrate the effectiveness of the proposed framework and demonstrate that the super-resolved fields match the accuracy of an advanced numerical solver running at 400 times the coarse mesh resolution while simultaneously satisfying the (highly nonlinear) governing laws. The approach opens the door to applying machine learning and traditional numerical approaches in tandem to reduce  computational complexity accelerate scientific discovery and engineering design.

\blfootnote{Accepted at SC22: Workshop on Art\mbox{}if\mbox{}icial Intelligence and Machine Learning for Scientif\mbox{}ic Applications}
\end{abstract}

\pagenumbering{arabic}



\section*{Impact Statement}
 Accurate modeling of macroscale behavior of complex heterogeneous materials is important for industrial applications but is also computationally expensive. The physics-informed deep-learning based super-resolution (PhySRNet) framework we introduce in this paper is aimed at overcoming this computational challenge. PhySRNet enables  researchers to run their numerical simulations on a coarse mesh and allows them to successfully reconstruct super-resolved ($400$ times coarse mesh resolution) solution fields from the low-resolution coarse mesh simulation results while achieving great accuracy. We believe the proposed framework provides possibilities for guiding future subgrid-scale models for modeling complex phenomena occurring at small spatial and temporal scales. This machine learning accelerated approach to material modeling will enable computationally efficient modeling and fast discovery of improved materials that exhibit desired combination of mechanical and thermal properties which would have a large impact on industries as diverse as biomedical, automotive, infrastructure, and electronics.
 



\section{Introduction}
\label{sec:introduction}




Numerical methods such as Finite element method \cite{hughes2012finite}, Isogeomteric analysis \cite{cottrell2009isogeometric},  and mesh-free methods \cite{liu1995reproducing, belytschko1994element} are few of the conventional approaches employed in solving the Partial Differential Equations (PDEs) involved in computational solid mechanics problems. However, the ever-increasing sophistication of material models by incorporating more complex physics, such as modeling  size-effect  \cite{fleck1994strain, arora2020unification} or dislocation density evolution \cite{arora2020finite, arora2019computational, arora2020dislocation, arora2022mechanics, joshi2020equilibrium}, or advanced materials such as composites and multicomponent alloys with spatially-varying material properties (heterogeneity) and direction dependent behavior (anisotropy) is bringing these numerical solvers to their limits. Hence, it is becoming a formidable task to perform simulations that can resolve  the complex physical phenomena occurring at small spatial and temporal scales and  accurately predict the macro-scale behavior of materials. Therefore, a cost-effective physics-based surrogate model that allows the researchers to perform simulations on a coarse mesh without sacrificing accuracy will be highly beneficial for many reasons. \underline{First}, researchers can choose to run their simulations at a lower resolution (\textit{online} stage) and later reconstruct the solution back to the target resolution (\textit{offline} stage). This will significantly reduce the computational expense during the \textit{online} stage,  thus accelerating the process of scientific investigation and discovery. \underline{Second}, the surrogate model based on data super-resolution can also be used to enhance outputs from experimental techniques for full-field displacement and strain measurement such as Digital Image Correlation (DIC) which would allow researchers to generate and store a small fraction of data.

Recent advances in Deep Learning (DL) and Physics-Informed Neural Networks (PINN) \cite{raissi2017physicsI, raissi2019physics} make it a promising tool to tackle this computational challenge.  Several applications of PINNs can be found in the literature ranging from modeling of fluid flows and Navier Stokes equations \cite{sun2020surrogate,rao2020physics,jin2021nsfnets}, cardiovascular systems \cite{kissas2020machine,sahli2020physics}, and material modeling  \cite{frankel_prediction_2020, arora2022physics, shrivastava2022predicting, zhu2021machine}, among others.  More recently, inspired by the growing success of image super-resolution techniques in the field of computer vision \cite{dong2014learning,lai2017deep, haris2019recurrent, zhang2018residual},  researchers have explored the possibility of using deep learning based super-resolution (SR) technique to reconstruct high-resolution (HR) fluid flow fields from low-resolution (LR) (possibly noisy) data \cite{fukami2021machine,fukami2019super,deng2019super,bode2019using,xie2018tempogan,    esmaeilzadeh2020meshfreeflownet,subramaniam2020turbulence,sun2020physics,gao2021super}.  In a proof of concept, Arora \cite{arora2021machine} investigated the application of physics-informed SR for a linear elasticity problem. However, technical challenges still remain in developing a physics-informed DL based model for super-resolution in computational solid mechanics in label-free scenario, especially for  materials undergoing large deformation. 

While the data-driven approaches for reconstructing HR flow fields  have also shown promising results, these approaches require large amount of computationally expensive HR labeled data for training. Moreover, the output solution fields may fail to satisfy the governing laws of the system  (PDEs and initial/boundary conditions) since these models lack any physics-based constraints.





In this paper,  we propose a physics-informed deep learning based super-resolution framework (PhySRNet) mechanics without requiring any HR labeled data. In particular, we explore and demonstrate the effectiveness of PhySRNet for resolving the LR displacement and stress fields in the body undergoing hyperelastic deformation in the absence of any HR data. The LR input data is obtained by running simulations on a coarse mesh which is $400$ times coarser than the target resolution. Furthermore, the chosen material model also presents a special scenario wherein the  model initialization plays an important role in guiding its convergence which makes the application to nonlinear solids significantly distinguishing from prior works involving super-resolution of fluid flow.

The layout of the rest of this paper is as follows: In Sec.~\ref{sec:background}, a brief review of the governing equations for modeling hyperelastic deformation in solids is presented. Model architecture and construction of physics-based loss function are discussed in sections \ref{sec:architecture}  and \ref{sec:loss}, respectively. The simulation setup for generating LR displacement and stress field data, to be used as input, for training and evaluating the machine learning framework is presented in Sec.~\ref{sec:data_collection}. Sec.~\ref{sec:results} presents the results that demonstrate the effectiveness of the proposed framework in super-resolving the stress and displacement fields for the example problem  considered. Conclusions and future opportunities are presented in Sec.~\ref{sec:conclusion}.

%

\section{Background}
\label{sec:background}
This section presents the governing equations for modeling the hyperelastic behavior in solids.

\subsection{Governing equations for hyper-elastic modeling}
\label{sec:equations}
We briefly recall the governing equations for modeling the behavior of hyperelastic solids. The reader is referred to standard textbooks  \cite{gurtin_fried_anand_2010} for a detailed discussion on the thermodynamics and mechanics of continuous media. We use the mixed-variable formulation, i.e., displacement vector and stress tensor fields $(\bm{u}, \bm{P})$ as unknowns  in this work. This formulation is shown to be of  crucial importance to ensure greater numerical accuracy of the solution and avoiding convergence issues for linear \cite{rao2021physics} and nonlinear \cite{arora2022physics} cases during model training. The governing  equations, in the absence of inertial forces, are given as follows:
\begin{align}
	\begin{split}
		Div{\bm{P}} & + \bm{b} = \bm{0}, ~~\text{in}~~\Omega,\\
		\bm{P} \bm{N} = \bm{t}_{bc} ~~ \text{on}~~ & \partial\Omega_{N} \text{~~and~~}	\bm{u} = \bm{u}_{bc}~~ \text{on}~~\partial\Omega_{D}.
	\end{split}
	\label{eq:sys1}
\end{align}
In the above, $\bm{P}$ denotes the first Piola-Kirchhoff stress, and $\bm{b}$ denotes the body force per unit volume of the undeformed (reference) configuration $\Omega$ of the material. $Div$ denotes the divergence operator in the configuration $\Omega$.  $\bm{t}_{bc}$ and $\bm{u}_{bc}$ denote the known traction and displacement vectors on the (non-overlapping) parts of the boundary $\partial\Omega_{N}$ and $\partial\Omega_{D}$, respectively. $\bm{N}$ denotes the unit outward normal to the external boundary $\partial\Omega$. 

In this work, the material is assumed to behave as a compressible analog of Neo-Hookean material whose stored energy density and constitutive relationship are given as 
\begin{align}
    \psi = \frac{\mu}{2} \, &\left(\textrm{trace}(\bm{B})-3\right) - \mu \ln{J} ~~;~~ J = \det(\bm{F}),\label{eq:energy}\\
    &\quad \bm{P} = \mu\left( \bm{B}-\bm{I}\right),
    \label{eq:piola}
\end{align}
where $\mu$ denotes the shear modulus of the material, and  $\bm{I}$ denotes the second order identity tensor. $\bm{B}=\bm{FF}^T$  denotes the left Cauchy–Green tensor  and  $\bm{F} = \bm{I} + \bm{\nabla}\bm{u}$ denotes the deformation gradient. We emphasize that finite deformation material modeling requires special consideration during the initialization of the model's weights  (discussed in section \ref{sec:loss}) since the constraint $J(X,Y)>0$ has to be strictly satisfied at every point $(X,Y)$ in the undeformed configuration.  Under two-dimensional plane-strain conditions, the unknown components for displacement vector $\bm{u}$ and stress tensor $\bm{P}$ are $(\bm{u}_x,\bm{u}_y)$ and $(\bm{P}_{xx}, \bm{P}_{yy}, \bm{P}_{xy}, \bm{P}_{yx})$, respectively.





\section{Methodology}
\label{sec:method}

This section discusses the architecture of  PhySRNet followed by the discussion on construction of the loss function for the training of the model.

\subsection{Model Architecture}
\label{sec:architecture}

\begin{figure}[t]
	\centering
	{\includegraphics[width=.98\linewidth]{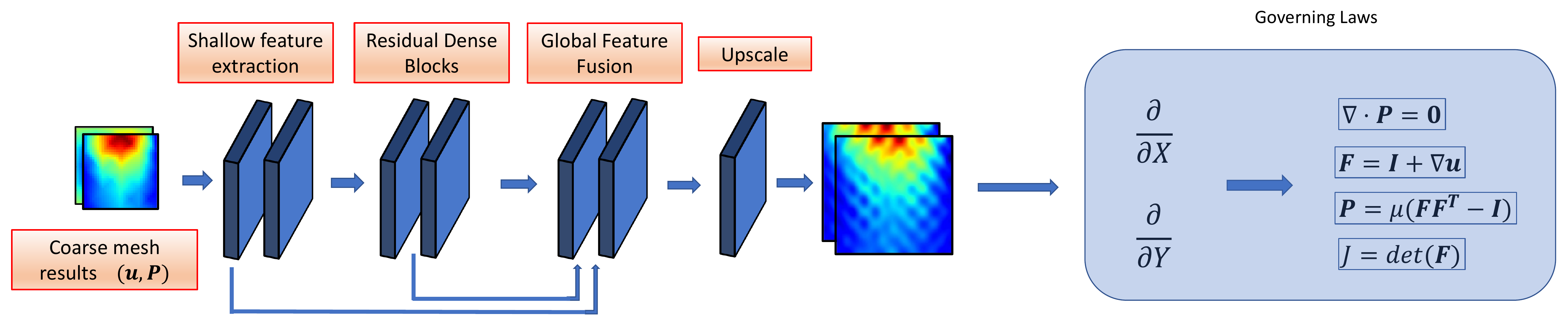}}
	\caption{The schematic of physics-informed super-resolution framework.}
	\label{fig:sch_sr}
\end{figure}

The composite architecture of the physics-informed super-resolution network (PhySRNet) proposed in this work is shown in Fig.~\ref{fig:sch_sr}. We use a separate neural network to resolve each solution field individually as this composite learning structure with decoupled sub-networks has been shown to be effective in enhancing the learning performance for multivariate regression problems \cite{guo2020ssr}. This approach also allows the trainable network parameters to be decoupled for each solution field with varying magnitudes and spatial distribution.







The architecture of individual sub-networks is built upon the Residual Dense Network (RDN) proposed in \cite{zhang2018residual}. The RDN architecture comprises four parts: a) shallow feature extraction net (SFENet), b) residual dense blocks (RDBs), c) dense feature fusion (DFF), and d) up-sampling net (UPNet). The architecture has several unique advantages when compared to other SR architectures \cite[sec.~4]{zhang2018residual} including the property to extract abundant local features via dense convolutional layers from LR input and the ability to adaptively fuse the hierarchical features in a global way.

\begin{figure}[htp]
	\centering
		\begin{subfigure}[b]{.250\linewidth}
		\centering
		{\includegraphics[width=0.995\linewidth]{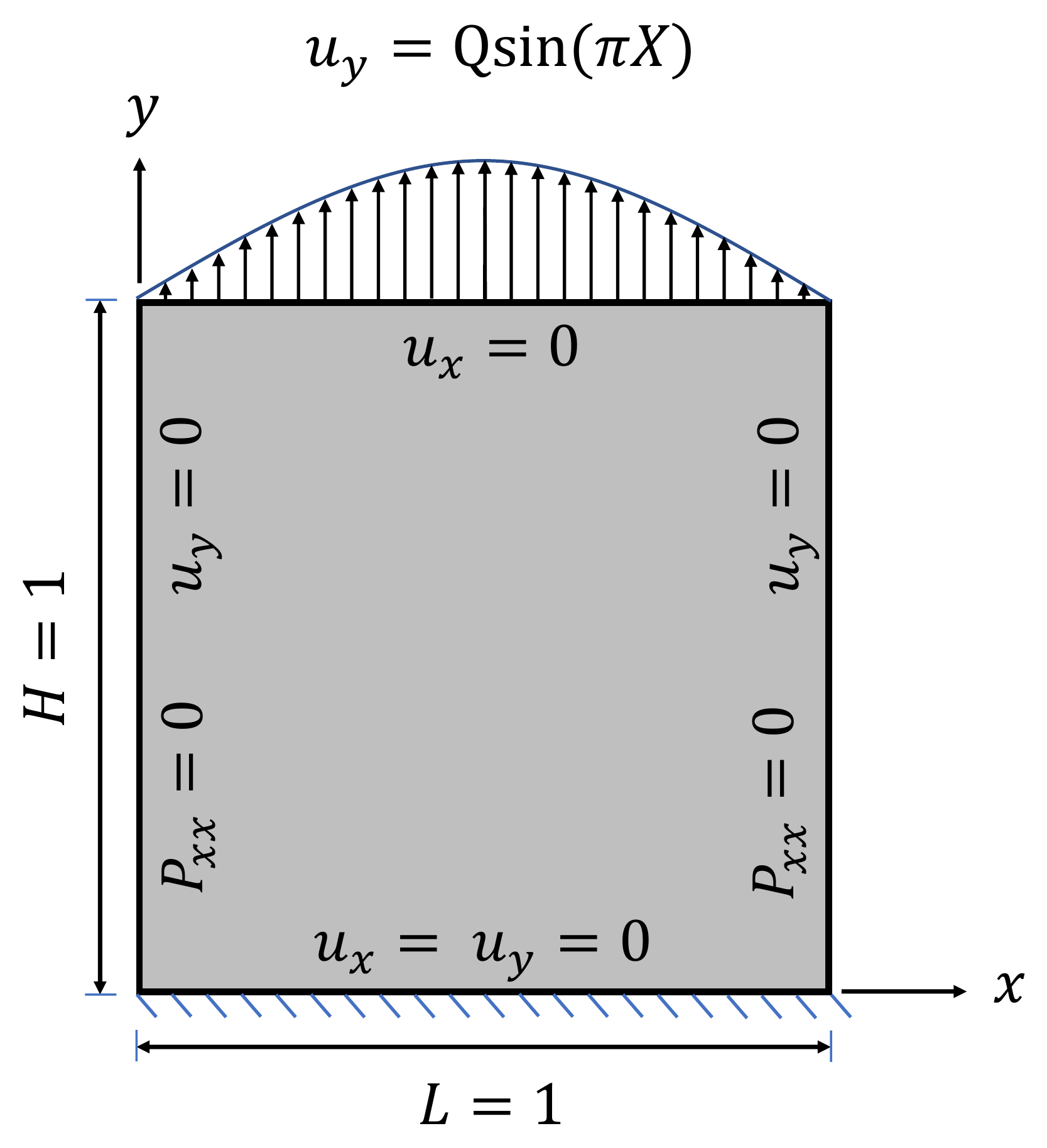}}
		\caption{}
		\label{fig:sch_body}
	\end{subfigure}%
		\begin{subfigure}[b]{.200\linewidth}
	\centering
	{\includegraphics[width=0.995\linewidth]{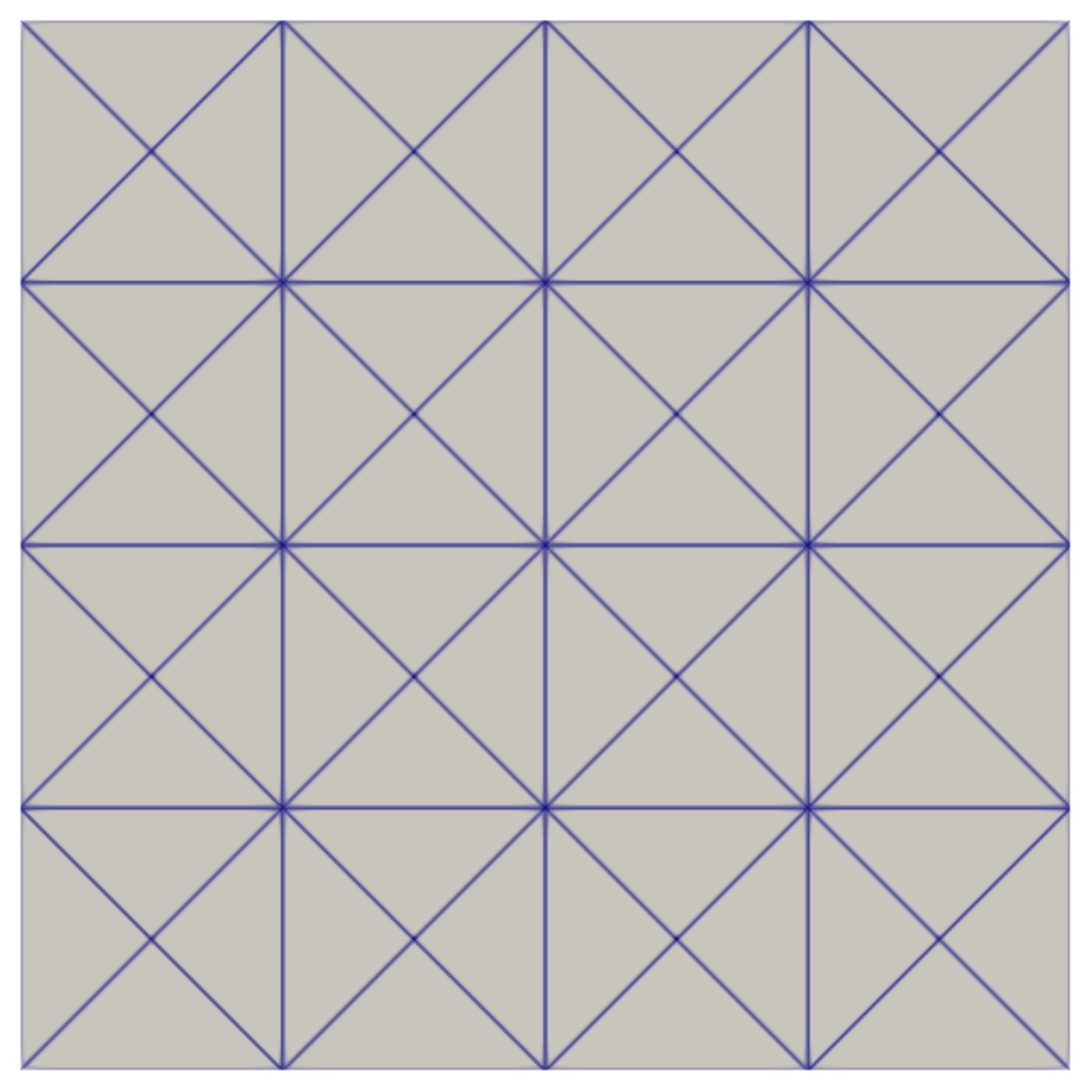}}
			\caption{}\label{fig:coarse_mesh}\vspace{10pt}
\end{subfigure}\hspace{20pt}
	\begin{subfigure}[b]{.200\linewidth}
	\centering
	{\includegraphics[width=0.995\linewidth]{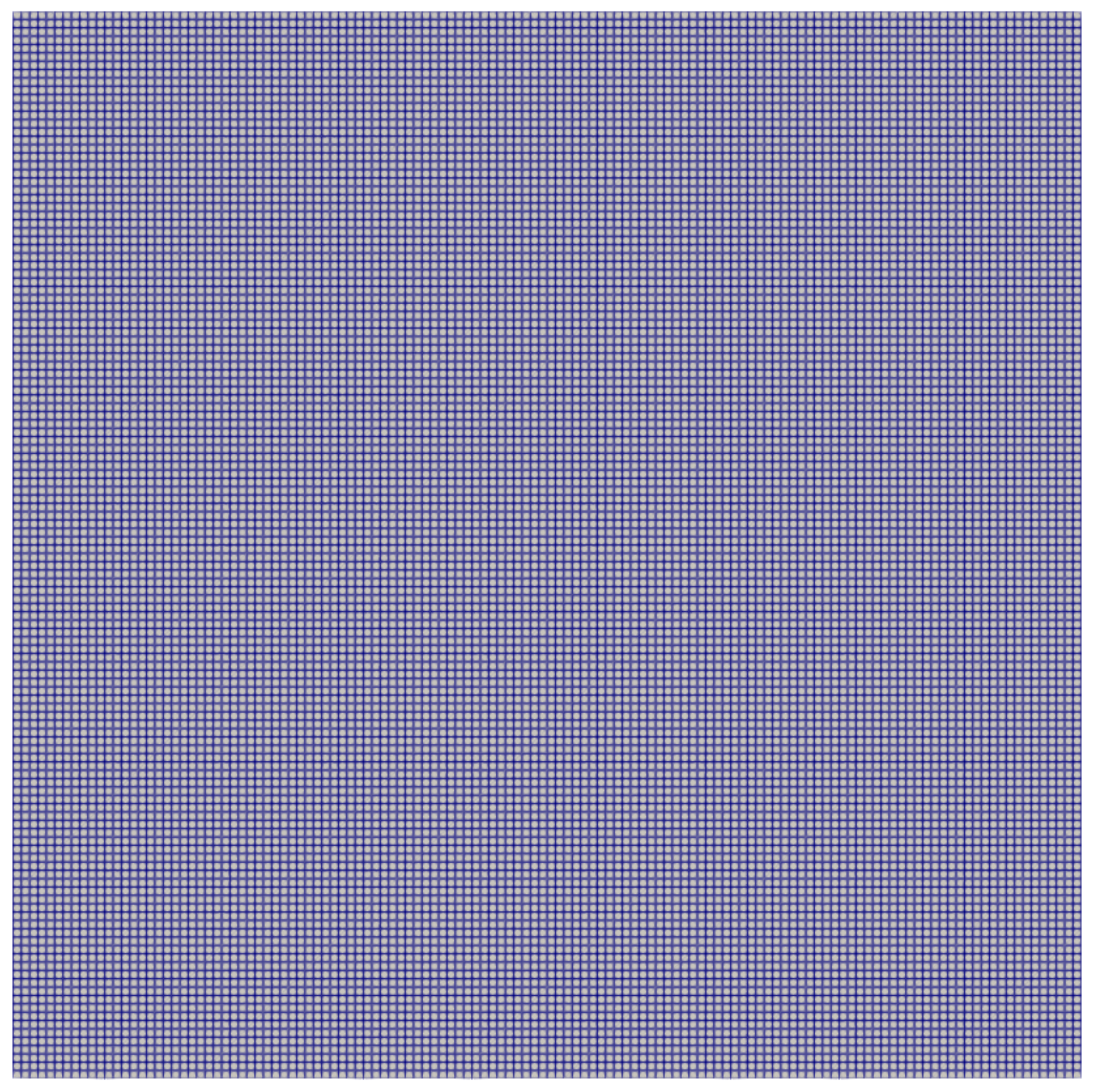}}		\caption{}\label{fig:fine_mesh}\vspace{10pt}
\end{subfigure}
	\caption{a) Schematic showing the geometry and the applied boundary conditions. b) Coarse triangular mesh with $41$ nodes. c) $128 \times 128$ fine mesh with $16384$ nodes. The LR data is refined by $\approx400$ times.}
	\label{fig:fig_army}
\end{figure}

Each sub-net uses the following hyper-parameters: number of residual blocks: $2$, number of layers in each residual block: $4$, growth rate: $32$, and number of features: $32$. The inputs to the model consist of LR data $(u_x, u_y, P_{xx}, P_{yy}, P_{xy}, P_{yx})$ obtained by running simulations on a coarse mesh (see Fig.~\ref{fig:coarse_mesh}) and then interpolating the solution (using FEM basis functions) on a $32 \times 32$ structured grid. The outputs of the framework correspond to the HR data on a $128 \times 128$ structured grid shown in Figure \ref{fig:fine_mesh}.





\subsection{Constructing the loss function}
\label{sec:loss}
In the absence of any HR labeled data, the network's total loss $\mathcal{L}$ is obtained from the physics-based constraints of the system - governing PDEs, constitutive law, and boundary/initial conditions. However, it has been well documented that presence of multiple components in $\mathcal{L}$ gives rise to competing effects amongst them leading to convergence issues during the training of the model \cite{bischof2021multi, wang2020understanding, arora2022physics}. Therefore, in this work, we choose to impose the boundary conditions in a ``hard" manner thus eliminating their contribution from total loss $\mathcal{L}$.




For each unknown field component, the boundary conditions are imposed in a hard manner by using a composite scheme which consists of using  a function $\mathcal{F}$ that satisfies the boundary condition, a function $\mathcal{G}$ that is zero on the Dirichlet boundaries, and the output $\mathcal{N}$ of the DL model. The final solution to the super-resolution problem for each output field $\Phi \in \{u_x, u_y, P_{xx}, P_{yy},  P_{xy}, P_{yx} \} $ is  then given as  follows:
\begin{align}
\begin{split}
		\Phi(X, Y) &= \mathcal{F}_{\Phi}(X, Y) + \mathcal{N}_{\Phi}(X, Y) \cdot \mathcal{G}_{\Phi}(X, Y)
\end{split}
\end{align}


\noindent  For complicated geometries and boundary conditions, the above strategy can be generalized to obtain functions $\mathcal{F}$ and $\mathcal{G}$ as outputs of separate DL models as shown in \cite{rao2021physics}. 
For finite deformation material modeling, the physical constraint $J(X,Y) > 0$  has to be strictly enforced as well. Since the boundary conditions are satisfied exactly, the total loss $\mathcal{L}$ is given as follows 
\begin{equation}
	\mathcal{L} = \lambda_1 \, \underbrace{L( Div\bm{P}, \,\bm{0})}_{\textrm{PDE}} + \lambda_2 \, \underbrace{L(\bm{P}, \,\mu\left( \bm{B} - \bm{I} \right))}_{\textrm{Constitutive law}} + \underbrace{\mathcal{L}_J}_{J > 0},
\end{equation}
where $L(P, Q)$ measures the mean absolute error (MAE) for the  prediction $P$ and target $Q$. The constraint loss $\mathcal{L}_J$ is given as $\mathcal{L}_J = || \min(0, det\bm{F}) ||_{1}$ where $||(\cdot)||_1$ denotes the $L^1$ norm of the quantity $(\cdot)$. We utilize fourth order finite difference scheme to evaluate the derivatives of the fields on the fine mesh. 

The current work involving super-resolution to nonlinear solid mechanics problems has a significantly distinguishing feature from its application to resolving fluid flow in that any technique for initialization of model weights would not satisfy the physical constraint $J(X,Y)>0$ initially. This leads to nonconvergence during model training using any optimization strategies such as the family of gradient descent methods. To remedy this, we first train the model to satisfy the constraint exactly ($\mathcal{L}_J = 0$) by setting $\lambda_1 = \lambda_2 = 0$ during the few initial epochs of the training. After that, we choose $\lambda_1 = 1$ and $\lambda_2 = 10$ based on the heuristics that the equilibrium equation \eqref{eq:sys1} also depends the derivatives of $\mu$ for heterogeneous materials.








%
%

The framework is implemented and trained using PyTorch framework \cite{NEURIPS2019_9015}. The network's total loss $\mathcal{L}$ is minimized by iteratively updating its trainable parameters. The whole training process consists of two stages: i) Initial convergence using Adam optimizer \cite{kingma_adam:_2015} with an initial learning rate $\eta=10^{-3}$, and ii) use of L-BFGS optimizer until the loss finally converges to a small value. While using Adam optimizer, the learning rate is also adaptively reduced by using \texttt{ReduceLROnPlateau} scheduler with the \texttt{patience}  set to $40$. The source code for the proposed framework along with the dataset used in this research can be found at \url{https://github.com/sairajat/SuperResolutionFiniteDeformation/} ~upon acceptance of this paper.


\begin{figure}[t]
	\centering
	\begin{subfigure}[b]{.120\linewidth}
		\centering
		\tiny \text{LR input}\par
		{\includegraphics[width=0.995\linewidth]{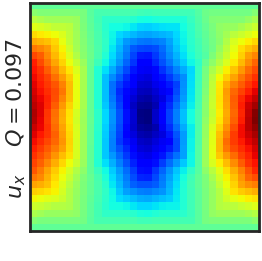}}
	\end{subfigure}%
	\begin{subfigure}[b]{.120\linewidth}
		\centering
		\tiny Bicubic\par
		{\includegraphics[width=0.995\linewidth]{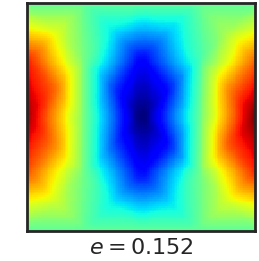}}
	\end{subfigure}%
	\begin{subfigure}[b]{.120\linewidth}
		\centering
		\tiny PhySRNet\par
		{\includegraphics[width=0.995\linewidth]{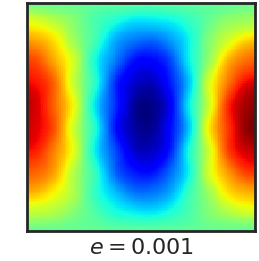}}
	\end{subfigure}%
	\begin{subfigure}[b]{.120\linewidth}
		\centering
		\tiny HR reference\par
		{\includegraphics[width=0.995\linewidth]{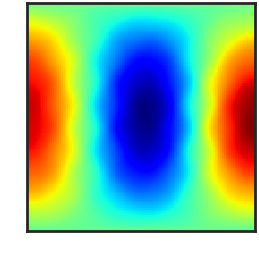}}
	\end{subfigure}\hfill
	\begin{subfigure}[b]{.120\linewidth}
		\centering
		\tiny LR input \par
		{\includegraphics[width=0.995\linewidth]{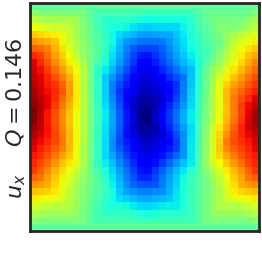}}
	\end{subfigure}%
	\begin{subfigure}[b]{.120\linewidth}
		\centering
		\tiny Bicubic\par
		{\includegraphics[width=0.995\linewidth]{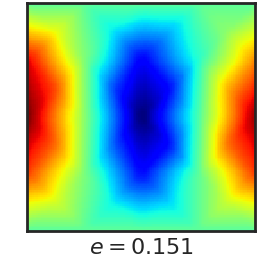}}
	\end{subfigure}%
	\begin{subfigure}[b]{.120\linewidth}
		\centering
		\tiny PhySRNet\par
		{\includegraphics[width=0.995\linewidth]{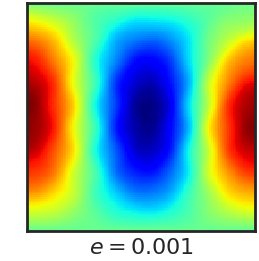}}
	\end{subfigure}%
	\begin{subfigure}[b]{.120\linewidth}
		\centering
		\tiny HR reference\par
		{\includegraphics[width=0.995\linewidth]{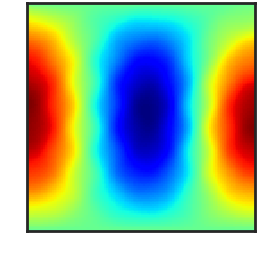}}
	\end{subfigure}\\[0.20em]

	\begin{subfigure}[b]{.120\linewidth}
		\centering
		{\includegraphics[width=0.995\linewidth]{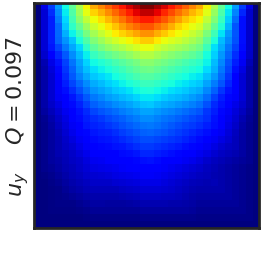}}
	\end{subfigure}%
	\begin{subfigure}[b]{.120\linewidth}
		\centering
		{\includegraphics[width=0.995\linewidth]{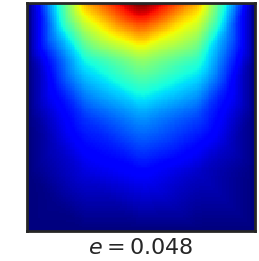}}
	\end{subfigure}%
	\begin{subfigure}[b]{.120\linewidth}
		\centering
		{\includegraphics[width=0.995\linewidth]{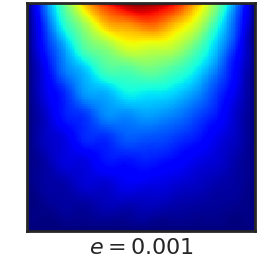}}
	\end{subfigure}%
	\begin{subfigure}[b]{.120\linewidth}
		\centering
		{\includegraphics[width=0.995\linewidth]{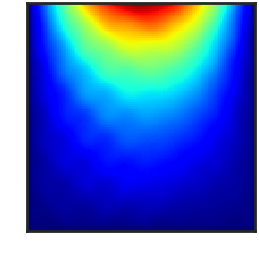}}
	\end{subfigure}\hfill
	\begin{subfigure}[b]{.120\linewidth}
		\centering
		{\includegraphics[width=0.995\linewidth]{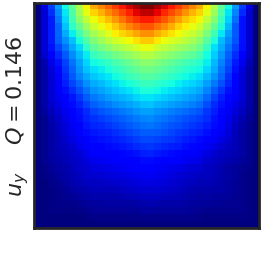}}
	\end{subfigure}%
	\begin{subfigure}[b]{.120\linewidth}
		\centering
		{\includegraphics[width=0.995\linewidth]{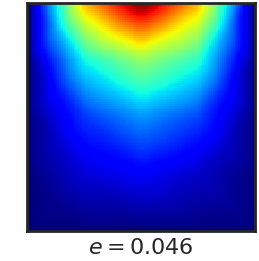}}
	\end{subfigure}%
	\begin{subfigure}[b]{.120\linewidth}
		\centering
		{\includegraphics[width=0.995\linewidth]{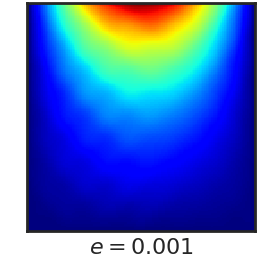}}
	\end{subfigure}%
	\begin{subfigure}[b]{.120\linewidth}
		\centering
		{\includegraphics[width=0.995\linewidth]{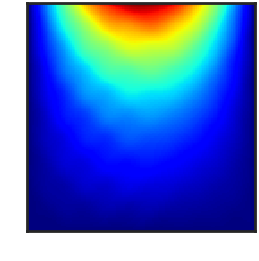}}
	\end{subfigure}\\[0.20em]

	\begin{subfigure}[b]{.120\linewidth}
		\centering
		{\includegraphics[width=0.995\linewidth]{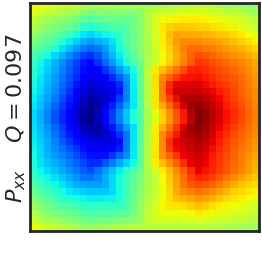}}
	\end{subfigure}%
	\begin{subfigure}[b]{.120\linewidth}
		\centering
		{\includegraphics[width=0.995\linewidth]{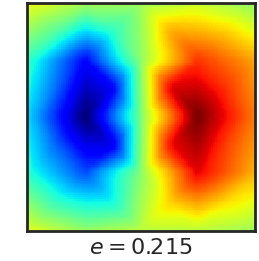}}
	\end{subfigure}%
	\begin{subfigure}[b]{.120\linewidth}
		\centering
		{\includegraphics[width=0.995\linewidth]{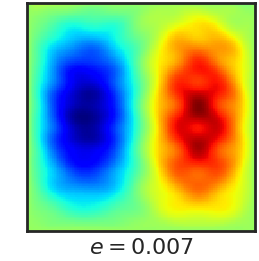}}
	\end{subfigure}%
	\begin{subfigure}[b]{.120\linewidth}
		\centering
		{\includegraphics[width=0.995\linewidth]{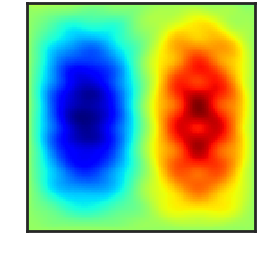}}
	\end{subfigure}\hfill
	\begin{subfigure}[b]{.120\linewidth}
		\centering
		{\includegraphics[width=0.995\linewidth]{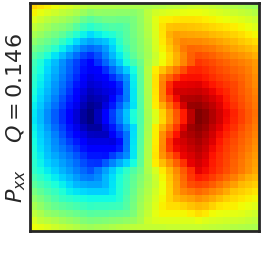}}
	\end{subfigure}%
	\begin{subfigure}[b]{.120\linewidth}
		\centering
		{\includegraphics[width=0.995\linewidth]{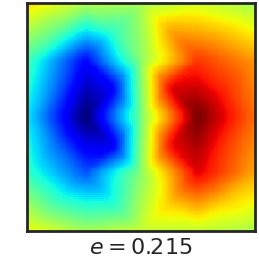}}
	\end{subfigure}%
	\begin{subfigure}[b]{.120\linewidth}
		\centering
		{\includegraphics[width=0.995\linewidth]{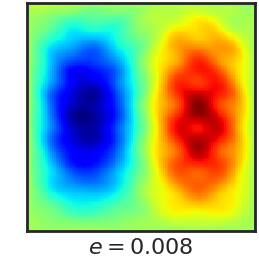}}
	\end{subfigure}%
	\begin{subfigure}[b]{.120\linewidth}
		\centering
		{\includegraphics[width=0.995\linewidth]{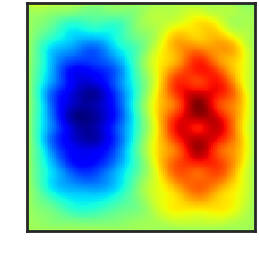}}
	\end{subfigure}\\[0.20em]

	\begin{subfigure}[b]{.120\linewidth}
		\centering
		{\includegraphics[width=0.995\linewidth]{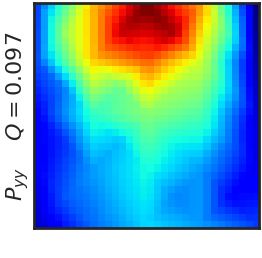}}
	\end{subfigure}%
	\begin{subfigure}[b]{.120\linewidth}
		\centering
		{\includegraphics[width=0.995\linewidth]{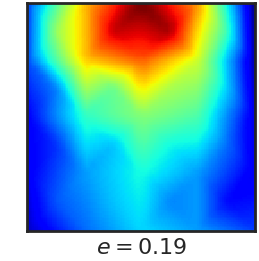}}
	\end{subfigure}%
	\begin{subfigure}[b]{.120\linewidth}
		\centering
		{\includegraphics[width=0.995\linewidth]{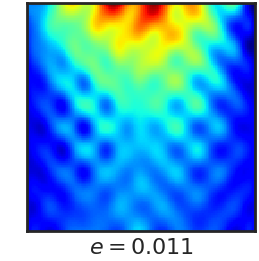}}
	\end{subfigure}%
	\begin{subfigure}[b]{.120\linewidth}
		\centering
		{\includegraphics[width=0.995\linewidth]{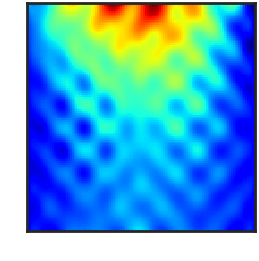}}
	\end{subfigure}\hfill
	\begin{subfigure}[b]{.120\linewidth}
		\centering
		{\includegraphics[width=0.995\linewidth]{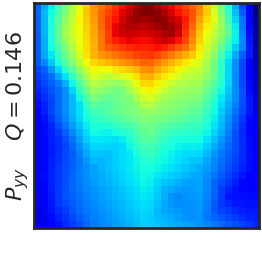}}
	\end{subfigure}%
	\begin{subfigure}[b]{.120\linewidth}
		\centering
		{\includegraphics[width=0.995\linewidth]{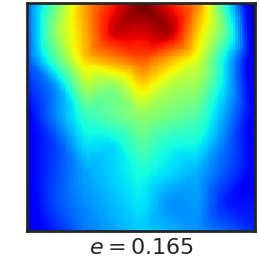}}
	\end{subfigure}%
	\begin{subfigure}[b]{.120\linewidth}
		\centering
		{\includegraphics[width=0.995\linewidth]{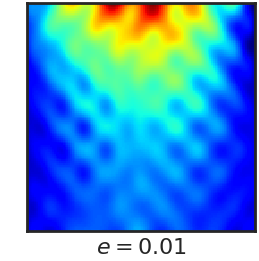}}
	\end{subfigure}%
	\begin{subfigure}[b]{.120\linewidth}
		\centering
		{\includegraphics[width=0.995\linewidth]{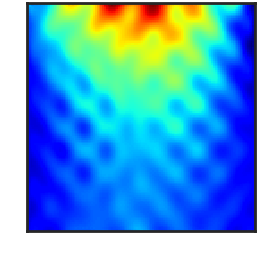}}
	\end{subfigure}\\[0.20em]

	\begin{subfigure}[b]{.120\linewidth}
		\centering
		{\includegraphics[width=0.995\linewidth]{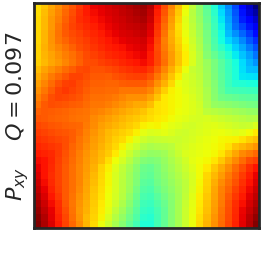}}
	\end{subfigure}%
	\begin{subfigure}[b]{.120\linewidth}
		\centering
		{\includegraphics[width=0.995\linewidth]{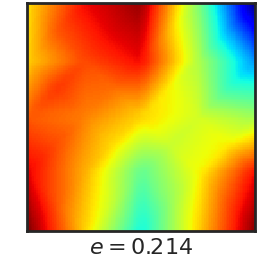}}
	\end{subfigure}%
	\begin{subfigure}[b]{.120\linewidth}
		\centering
		{\includegraphics[width=0.995\linewidth]{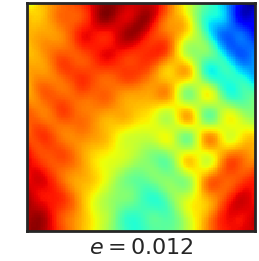}}
	\end{subfigure}%
	\begin{subfigure}[b]{.120\linewidth}
		\centering
		{\includegraphics[width=0.995\linewidth]{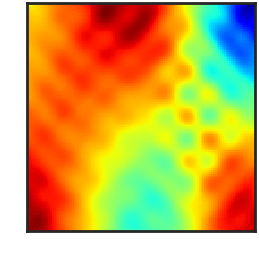}}
	\end{subfigure}\hfill
	\begin{subfigure}[b]{.120\linewidth}
		\centering
		{\includegraphics[width=0.995\linewidth]{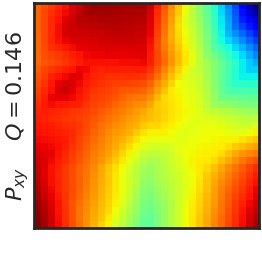}}
	\end{subfigure}%
	\begin{subfigure}[b]{.120\linewidth}
		\centering
		{\includegraphics[width=0.995\linewidth]{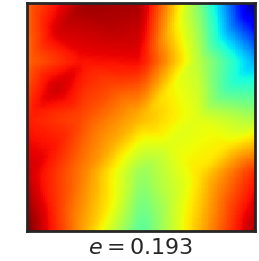}}
	\end{subfigure}%
	\begin{subfigure}[b]{.120\linewidth}
		\centering
		{\includegraphics[width=0.995\linewidth]{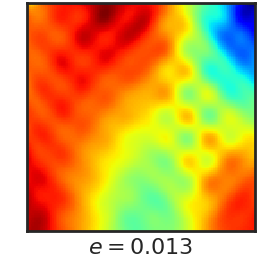}}
	\end{subfigure}%
	\begin{subfigure}[b]{.120\linewidth}
		\centering
		{\includegraphics[width=0.995\linewidth]{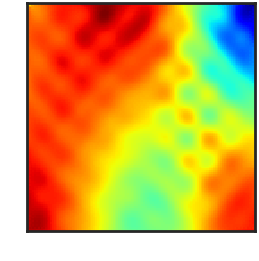}}
	\end{subfigure}\\[0.20em]

	\begin{subfigure}[b]{.120\linewidth}
		\centering
		{\includegraphics[width=0.995\linewidth]{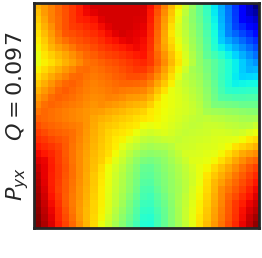}}
	\end{subfigure}%
	\begin{subfigure}[b]{.120\linewidth}
		\centering
		{\includegraphics[width=0.995\linewidth]{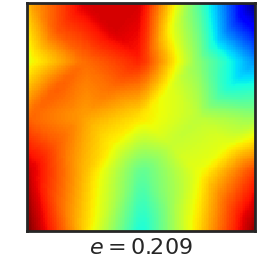}}
	\end{subfigure}%
	\begin{subfigure}[b]{.120\linewidth}
		\centering
		{\includegraphics[width=0.995\linewidth]{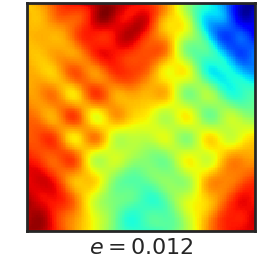}}
	\end{subfigure}%
	\begin{subfigure}[b]{.120\linewidth}
		\centering
		{\includegraphics[width=0.995\linewidth]{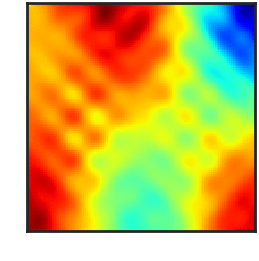}}
	\end{subfigure}\hfill
	\begin{subfigure}[b]{.120\linewidth}
		\centering
		{\includegraphics[width=0.995\linewidth]{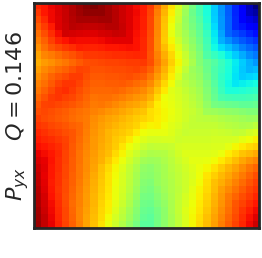}}
	\end{subfigure}%
	\begin{subfigure}[b]{.120\linewidth}
		\centering
		{\includegraphics[width=0.995\linewidth]{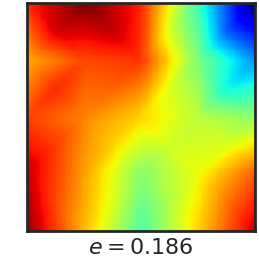}}
	\end{subfigure}%
	\begin{subfigure}[b]{.120\linewidth}
		\centering
		{\includegraphics[width=0.995\linewidth]{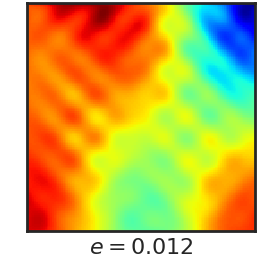}}
	\end{subfigure}%
	\begin{subfigure}[b]{.120\linewidth}
		\centering
		{\includegraphics[width=0.995\linewidth]{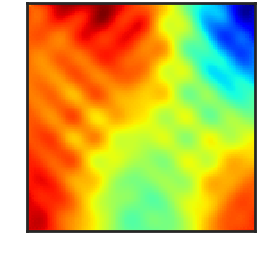}}
	\end{subfigure}
	\caption{The color contours of displacement vector and stress tensor components in two-dimensional elastic deformation 
		reconstructed with physics-informed super-resolution frameworks. Values below the plots indicate the  error  $e$. In both the blocks, the LR input data, HR ground truth data, bicubic interpolation, and the RDN output are
		plotted from the left to  right.}
	\label{fig:results}
\end{figure}

\section{Results \& Discussion}
\label{sec:results}




\subsection{Synthesis of low-resolution data}
\label{sec:data_collection}
To illustrate the application of the proposed approach, an example problem is setup as follows: We consider an isotropic body deforming quasi-statically under plane strain conditions subjected to the loading boundary conditions as shown in Fig.~\ref{fig:sch_body}. The body force vector $\bm{b} = (b_x,b_y)$ is given as 
\begin{align*}
	b_x &= \mu_s \left[ 9\pi^2 \cos(2\pi X) \sin(\pi Y) - \pi \cos(\pi X) Q Y^3 \right],\\
	b_y &= \mu_s \left[ -6 \sin(\pi X) Q Y^2 + 2\pi^2 \sin(2\pi X) \cos(\pi Y) 
	+ \, 0.25 \pi^2 \sin(\pi X) Q Y^4 \right],
\end{align*}
where $\mu_s$ is taken to be $0.50$. 
The shear modulus $\mu(X, Y)$ of the material is taken to be 
\begin{equation}
	\mu = \frac{\mu_s}{2} \left[3 + \sin(2\pi k X) \sin(2 \pi kY) \right];~~ k = 5,
\end{equation}
to represent a body with heterogeneous material properties. The scalar $Q \in [0.02, 0.20]$ directly affects the magnitudes of boundary conditions (see Fig.~\ref{fig:sch_body}) and the body force $\bm{b}$.  In this work, we use the following analytical forms of the functions $\mathcal{F}$ and $\mathcal{G}$ that ensure satisfaction of the boundary conditions 
\begin{align*}
	\mathcal{G}_{P_{xx}} = X\,(1-X)~~;~~
	\mathcal{G}_{u_x} = Y\,(1-Y)~~;~~
	\mathcal{G}_{u_y} = XY(1-X)(1-Y).
\end{align*}
The ground truth data is generated by solving the system of equations \eqref{eq:sys1} on a coarse mesh (shown in Fig.~\ref{fig:fig_army}) using Finite Element Method in Fenics \cite{alnaes2015fenics} for $100$ regularly sampled values of $Q$. The data is then randomly split in a $80:20$ ratio for training and test purposes.

\subsection{Application to hyperelasticity}
We now demonstrate the effectiveness of  PhySRNet by applying it to reconstruct the HR displacement and stress fields for the problem setup discussed  in section \ref{sec:data_collection}. The framework takes the coarse mesh solution fields interpolated to a $32\times32$ grid as inputs and outputs the solution fields on a $128\times128$ structured grid which is approximately $\!400$ times finer than the coarse mesh. We note that the framework presented herein can be generalized to work with non-rectangular domains by utilizing elliptic coordinate transformation as outlined in \cite{gao2020phygeonet}.

Figure \ref{fig:results} presents the results for the reconstructed displacement and stress fields obtained from PhySRNet for $2$ different values of $Q$. A simple bicubic interpolation of the solution fields and the HR reference data are plotted for comparison.  We note that the HR reference data is used only for the comparison with the model outputs. The figure clearly show that the reconstructed solutions fields are in great agreement with the HR reference data.  The model is successfully able to  resolve the spatial variation in the output fields  even though the LR inputs lacked such variation.    To quantitatively measure the accuracy, we define an  error measure $e$ as 
 \begin{equation}
     e = \frac{||\mathcal{I}^{HR} - \hat{\mathcal{I}}^{HR}||_{L^2}}{||\mathcal{I}^{HR}||_{L^2}},
 \end{equation} where $\hat{\mathcal{I}}^{HR}$ denotes the framework predictions.  The value of $e$ is reported underneath the reconstructed fields obtained using the PhySRNet and the bicubic interpolation. As can be seen from  figure \ref{fig:results}, the error $e$ is larger for data obtained from bicubic interpolation method since the outputs may not faithfully satisfy the  governing laws of the system.  The small values of $e$ for the model predictions signify that the reconstructed HR outputs obtained from PhySRNet almost match the accuracy of an advanced numerical solver running at  $400$ times the coarse mesh resolution. Therefore, we can  conclude that PhySRNet successfully enhanced the spatial resolution of the solution fields while ensuring that they satisfy the governing laws of the system.
 



\section{Conclusion \& future work}
\label{sec:conclusion}
In summary, we successfully trained and evaluated a physics-informed deep learning based super-resolution framework (PhySRNet) to reconstruct the deformation fields in a heterogeneous body undergoing hyperelastic deformation without requiring any HR labeled data. The approach is successfully able to learn high-resolution spatial variation of displacement and stress fields from their low-resolution counterparts for the example problem discussed. We show that the outputs from the PhySRNet match the accuracy of an advanced numerical solver running at $400$ times the coarse mesh resolution (see Figs.~\ref{fig:coarse_mesh} and \ref{fig:fine_mesh}). This approach exemplifies how machine-learning can be leveraged alongside numerical simulations to reduce the computational complexity and accelerate scientific discovery and engineering design without sacrificing accuracy.


While the current work focuses on nonlinear quasi-static problems, the future work aims to extend the framework for both spatial and temporal super-resolution of (unsteady) elastodynamics problems in two and three dimensions. Moreover, we also aim to modify the architecture of the PhySRNet to explore if a sequence of low-resolution data inputs could help to further improve the quality of the reconstruction. 




%

\section*{Acknowledgements}
This work was conceptualized during the author’s time at Carnegie Mellon University (CMU). It is a pleasure to acknowledge discussions with Prof.~Amit Acharya from CMU. The author also thank Ankit Shrivastava, research associate at Sandia National Laboratories, for useful discussions and comments on the manuscript.

\clearpage
\newpage

\bibliographystyle{alpha}
\bibliography{Arora_PhySRNet}

\newpage

\color{red}
{



}

\end{document}